\documentclass[showpacs, groupedaddress]{revtex4}
\usepackage[pdftex]{graphicx}

\begin{document}
\title{Geometric phases without geometry}

\author{Amar Vutha}
  \email{amar.vutha@yale.edu}
\author{David DeMille}
\affiliation{Department of Physics, Yale University, New Haven, CT 06520, USA}

\begin{abstract}
Geometric phases arise in a number of physical situations and often lead to systematic shifts in frequencies or phases measured in precision experiments. We describe, by working through some simple examples, a method to calculate geometric phases that relies only on standard quantum mechanical calculations of energy level shifts. This connection between geometric phases and off-resonant energy shifts simplifies calculation of the effect in certain situations. More generally, it also provides a way to understand and calculate geometric phases without recourse to topology.
\end{abstract}

\pacs{03.65.Vf, 
      32.60.+i  
      }

\maketitle

\section{Introduction}
A geometric phase, often also referred to as an adiabatic phase or Berry's phase, is a real, physical phase shift that can lead to measurable effects in experiments (\cite{firstexp,firstexp2,commins} are some examples). In the classic version of this effect, the state of a particle with a magnetic moment is modified under the influence of a magnetic field which undergoes a slow change in its direction. The motion of the field leads to a phase shift accumulated between the quantum states of the particle, in addition to the dynamical phase due to the Larmor precession of the magnetic moment around the magnetic field. This extra phase is termed a geometric phase because it has a simple interpretation in terms of the geometry traced out by the system's Hamiltonian as it evolves in its parameter space. We point the interested reader to \cite{holstein} and references within for an excellent account of the geometric approach to calculating geometric phases, and its various connections with the topology of the parameter space.

In practice however, the formulation of the geometric phase in terms of differential geometry can make its calculation somewhat cumbersome. For example, calculating geometric phases can be non-trivial when the path traced out in the parameter space is not a closed loop \cite{noncyclic}. In addition the geometric formulation is strictly correct only in the limit of adiabatic (slow) evolution, and corrections due to finite evolution speed can be important in some situations \cite{figure8}. Finally, the parameter space of the Hamiltonian can also become quite involved when the system is an atom or molecule with internal structure subject to various evolving fields (e.g. magnetic fields, electric fields, and their gradients). Far from being an abstract intellectual pursuit, our interest in calculating geometric phases for this sort of complicated system stems from precision atomic physics measurements, such as the search for permanent electric dipole moments \cite{FSB03}, where these phases can give rise to important systematic errors \cite{commins,pendlebury,lamoreaux} and/or quantum state decoherence \cite{shafer-ray, hinds}.

Energy level shifts from time-varying perturbations, such as Bloch-Siegert shifts and AC Stark shifts, are mostly treated in the literature as separate phenomena from geometric phases. However, the idea that both of these essentially involve the same physics is implicit in some recent work \cite{firstexp2, pendlebury, lamoreaux, cornellprivate} and originally derives from an analysis by Ramsey \cite{ramsey}. While most of these treatments analyze the simple case of a spin-1/2 particle in a time-dependent magnetic field, the link between geometric phases and off-resonant energy level shifts is in fact quite general. In this paper, we explore this connection by looking at some instructive examples, showing with simple perturbative calculations how energy level shifts lead to geometric phases in each case. We find that there are a number of advantages to this reformulation of the problem of calculating geometric phases. The energy shift approach to geometric phases can be applied to situations where the geometric formalism is opaque, and our calculations of energy shifts using perturbation theory only rely on standard undergraduate quantum mechanics. We feel that it is pedagogically useful to highlight the equivalence between geometric phases and the more familiar energy level shifts, thereby showing that there is nothing arcane about geometric phases.

In the following section we introduce our approach with the classic example of a spin-1/2 system subject to a magnetic field whose direction changes in time. Then in Section \ref{sec:two_wobbles} we work out modifications to the usual geometric phase when there are multiple Fourier components in the time evolution of the magnetic field. Sections \ref{sec:electric_field} and \ref{sec:EplusB} show how the geometric phase can arise in a system acted upon by electric fields or a combination of electric and magnetic fields. These sections also indicate how geometric phases can be analyzed in a system with a non-trivial level structure, such as an atom or molecule.

\section{Spin-1/2 system in a magnetic field}\label{sec:spin-1/2}

A spin $S$ system in a magnetic field is a well-studied example of the geometric phase \cite{firstexp, firstexp2,commins,holstein}. When the tip of the magnetic field vector slowly traces out a closed loop in space, the sublevels $|m_S\rangle$ of the system pick up an extra ``geometric'' phase $\phi_g(m_S) = m_S \times \Omega_s$ (in addition to the usual dynamical phase from Larmor precession) \cite{holstein,griffiths,berry,bkd}. Here $\Omega_s$ is the solid angle enclosed by the loop. In all the subsequent sections, without any loss of generality we focus on the observable quantity defined by the phase \emph{difference} between the $|m_S = \pm S\rangle$ sublevels: $\Delta \phi_g = \phi(m_S = +S) - \phi(m_S = -S)$. In this section we will examine a spin S = 1/2 system in a magnetic field and show how the geometric phase shift can be traced back to an off-resonant energy level shift, in this case an AC Zeeman (or Ramsey-Bloch-Siegert) shift.

\begin{figure}
\includegraphics[width=0.6\columnwidth]{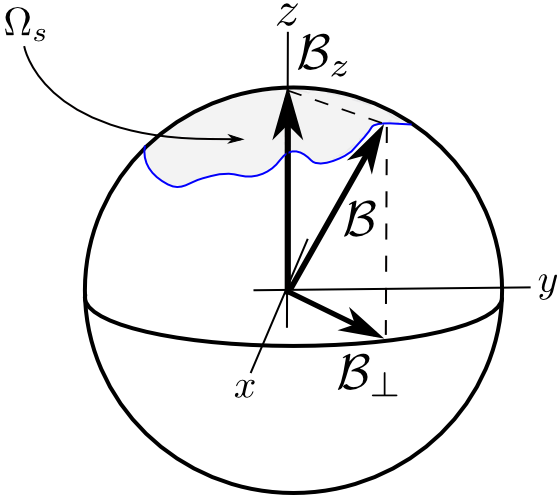}
\caption{An example of a trajectory traced by the magnetic field, showing how it can be decomposed into its components along a fixed set of axes: $\vec{\mathcal{B}} = \mathcal{B}_z \hat{z} + \vec{\mathcal{B}}_\perp(t)$. The geometric phase $\phi_g$ picked up between the sublevels of a spin-1/2 system interacting with this field is equal to the solid angle $\Omega_s$ enclosed by the trajectory.} 
\label{fig:solidangle}
\end{figure}

The usual approach to calculating the geometric phase tracks the instantaneous eigenstates of the system under the assumption of adiabatic evolution \cite{holstein,griffiths,berry,bkd}. In contrast, we shall use a fixed coordinate system and resolve the changing magnetic field into a static longitudinal ($\hat{z}$-directed) component and a dynamic transverse component (in the $xy$ plane), as shown in Figure \ref{fig:solidangle}. Let the evolution of the transverse component's direction be composed of a single angular frequency $\omega_\perp$. For a counter-clockwise rotation, the magnetic field written in terms of its components in the fixed coordinate system is 
\begin{equation}
\vec{\mathcal{B}} = \mathcal{B}_z \hat{z} + \vec{\mathcal{B}}_\perp(t)
\end{equation}
where the rotating transverse component of the magnetic field is
\begin{equation}
\vec{\mathcal{B}}_\perp(t) = \mathcal{B}_\perp (\hat{x} \ \cos \omega_\perp t + \hat{y} \ \sin \omega_\perp t).
\end{equation}
For this discussion we consider small values of $\mathcal{B}_\perp/\mathcal{B}_z$, such that the solid angle enclosed by this loop is $\Omega_s \approx \pi \frac{\mathcal{B}_{\perp}^2}{\mathcal{B}_z^2}$. 

\begin{figure}
\includegraphics[width=0.8\columnwidth]{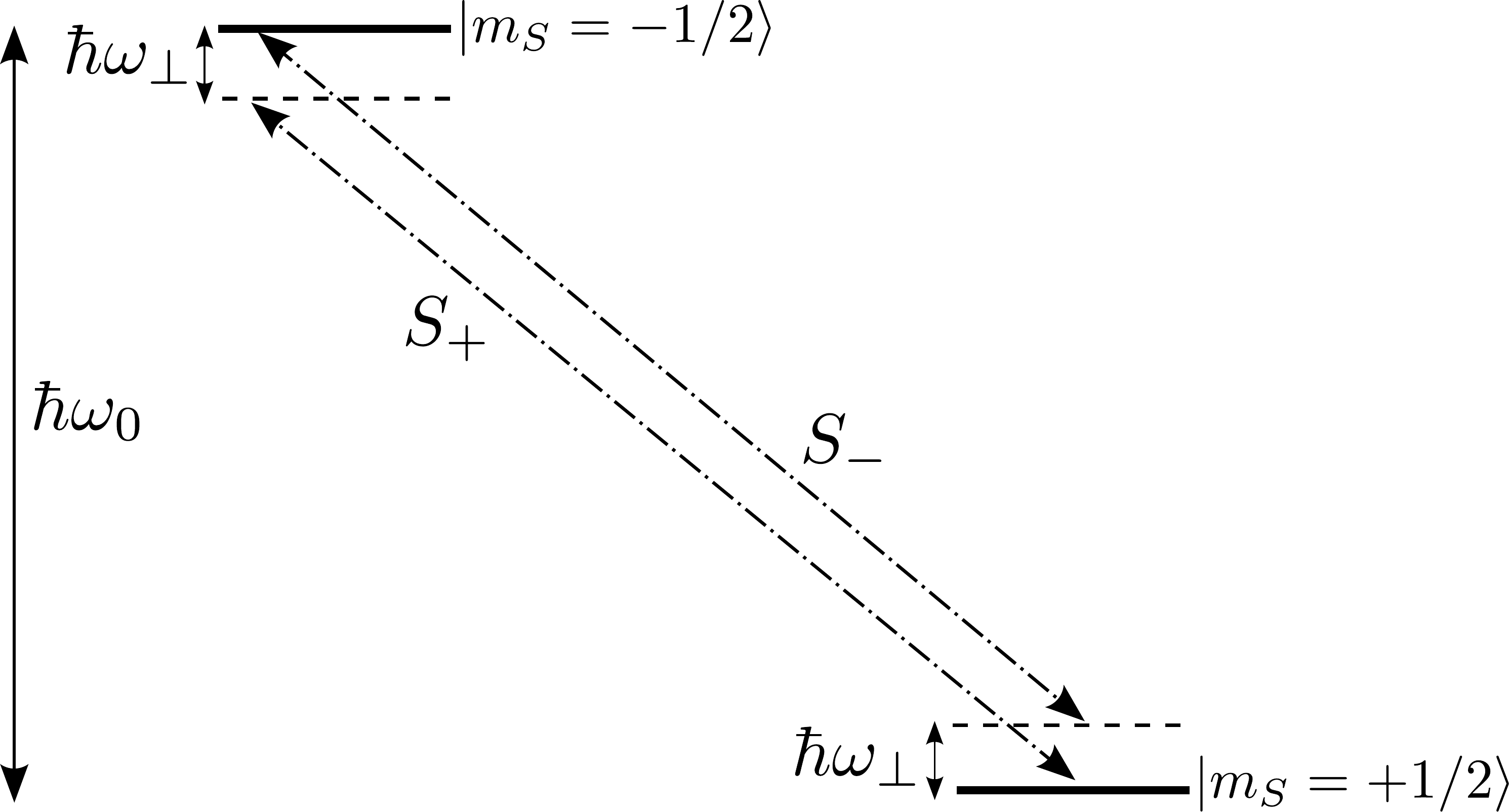}
\caption{Energy levels of a spin $S = 1/2$ system interacting with a magnetic field whose direction evolves over time. The level splitting in the longitudinal field is $\hbar \omega_0 = \gamma \mathcal{B}_z$. The dashed lines indicate virtual levels, offset from the real levels by the finite rotation frequency of the transverse field $\vec{\mathcal{B}}_\perp$. $S_\pm$ is the spin raising (lowering) operator which couples the $|m_S = \pm 1/2\rangle$ state to the [virtual] $|m_S = \mp 1/2\rangle$ state in the presence of $\vec{\mathcal{B}}_\perp$.} 
\label{fig:neutron}
\end{figure}

The Hamiltonian of the particle in this magnetic field is
\begin{eqnarray}
H_{int} & = & -\gamma \vec{S} \cdot \vec{\mathcal{B}} \\
& = & - \gamma S_z \mathcal{B}_z - \frac{\gamma \mathcal{B}_\perp}{2} \Big( \ S_{-} e^{-i \omega_{\perp} t} + S_{+} e^{i \omega_\perp t} \ \Big)
\end{eqnarray}
where $\gamma$ is the gyromagnetic ratio and $S_\pm = S_x \pm i S_y$ are the spin raising and lowering operators respectively. In the presence of only the longitudinal field $\mathcal{B}_z \hat{z}$, the eigenstates $|m_S = \pm 1/2\rangle$ have energies $E_{m_S} = \langle m_S|-\gamma S_z \mathcal{B}_z|m_S \rangle = -\gamma \mathcal{B}_z m_S$ and the Larmor precession frequency is $\omega_0 = (E_{-1/2} - E_{+1/2})/\hbar = \gamma \mathcal{B}_z/\hbar$. We will consider the effect of the time-varying transverse field as a perturbation on the $S_z$ eigenstates. To lowest order, the energy shifts $\Delta E_{\pm 1/2}$ of the $|\pm 1/2\rangle$ states are given by second order time-dependent perturbation theory. Figure \ref{fig:neutron} shows the energy levels and the operators connecting them. We define the transverse matrix elements of the spin operator to simplify notation
\begin{equation}
s_\perp^2 = |\langle +1/2|S_+|-1/2 \rangle|^2 = |\langle -1/2|S_-|+1/2 \rangle|^2 = 1.
\end{equation}

Taking note of the usual selection rules on the matrix elements of the $S_\pm$ operators, we get
\begin{eqnarray}
\Delta E_{+1/2} & = & \frac{1}{4} \Big[ \frac{\gamma^2 s_\perp^2 \mathcal{B}_{\perp}^2}{- \hbar \omega_0 + \hbar \omega_{\perp}} \Big] \nonumber \\
 & = & - \frac{1}{4} \Big[ \frac{\gamma \mathcal{B}_{\perp}^2}{\mathcal{B}_z} + \frac{\mathcal{B}_{\perp}^2}{\mathcal{B}_z^2} \hbar \omega_{\perp} + \mathcal{O}(\omega_\perp^2) \Big] \\
\nonumber \\
\Delta E_{-1/2} & = & \frac{1}{4} \Big[ \frac{\gamma^2 s_\perp^2 \mathcal{B}_{\perp}^2}{\hbar \omega_0 - \hbar \omega_{\perp}} \Big] \nonumber \\
 & = & \frac{1}{4} \Big[ \frac{\gamma \mathcal{B}_{\perp}^2}{\mathcal{B}_z} + \frac{\mathcal{B}_{\perp}^2}{\mathcal{B}_z^2} \hbar \omega_{\perp} + \mathcal{O}(\omega_\perp^2) \Big].
\end{eqnarray}
Here we have retained terms up to lowest order in $\omega_\perp$, the field's evolution frequency. Note that the condition for applicability of this approximation \emph{viz.} $\omega_\perp \ll \omega_0$ is the same as the adiabatic criterion invoked in the standard approach to describing the geometric phase \cite{griffiths}. 

The extra energy difference $\Delta E$ between $|\pm 1/2\rangle$ due to the transverse rotating component of the field is therefore
\begin{eqnarray}
\Delta E & = & \Delta E_{+1/2} - \Delta E_{-1/2} \nonumber \\
& = & \{ - \gamma \frac{\mathcal{B}_{\perp}^2}{2 \mathcal{B}_z} \} + [- \frac{1}{2}\frac{\mathcal{B}_{\perp}^2}{\mathcal{B}_z^2} \hbar \omega_{\perp}] = \Delta E_{QZ} + \Delta E_g,
\end{eqnarray}
where $\Delta E_{QZ}~(\Delta E_g)$ corresponds to the term in curly (square) brackets.  Note that the term $\Delta E_{QZ}$ is nonzero even when the field's direction evolves infinitesimally slowly. This term is nothing other than the correction to the Zeeman splitting because the total magnetic field becomes larger on application of the transverse field component: here $\mathcal{B} = \sqrt{\mathcal{B}_z^2 + \mathcal{B}_\perp^2} \approx \mathcal{B}_z + \frac{\mathcal{B}_\perp^2}{2\mathcal{B}_z}$. The term $\Delta E_g$ is responsible for the geometric phase. This term, which vanishes when $\omega_{\perp} \to 0$, nevertheless adds a relative phase between $|\pm 1/2\rangle$ even in this limit. The relative phase picked up over a time interval $T$ is 
\begin{equation}
\Delta \phi_g = - \Delta E_g T/\hbar = \frac{1}{2}\frac{\mathcal{B}_{\perp}^2}{\mathcal{B}_z^2} \omega_{\perp} T. \label{eq:alpha}
\end{equation}

Over one full cycle of the field's evolution, $T = 2 \pi/\omega_\perp$ and we get the standard geometric phase result $\Delta \phi_g = \pi \frac{\mathcal{B}_{\perp}^2}{\mathcal{B}_z^2} \simeq \Omega_s$. The restriction to small solid angles here is equivalent to truncating the perturbation series at second order in $\mathcal{B}_\perp$. Results in the case of larger solid angles can be analytically written down using higher orders of perturbation theory, or alternatively using a dressed-state formalism that is valid for arbitrarily large solid angle \cite{cornellprivate, cohen-tannoudji}. However the essential point remains that energy level shifts can be calculated algebraically (by numerically diagonalizing the Hamiltonian if necessary), without having to track the eigenstates over a geometric path in parameter space.

This formulation can be easily extended to situations that are more complicated than the simple loop. Moreover, by keeping terms of higher order in $\omega_\perp$, corrections due to deviations from the adiabatic limit can be calculated. The energy shift formalism also eliminates the need to track the instantaneous quantization axis and basis states of the system throughout its evolution. The use of a fixed basis set makes contact with the standard tools of quantum mechanics as taught in most undergraduate courses. Having illustrated the basic idea, we now apply this formalism to a case where the evolution of the field's direction is more complicated.

\section{Spin-1/2 system with multiple evolution frequencies}\label{sec:two_wobbles}

\subsection{Geometric phase calculation using energy shifts}

Let us calculate the geometric phase in a situation where there are two Fourier components to the evolution of the magnetic field experienced by a spin $S=1/2$ system. This demonstrates the approach to be used in a general situation, \emph{e.g.} in a precision experiment where the magnetic field experienced by an atom along its trajectory can be quite complicated and contain a number of Fourier components. A magnetic field with two evolution frequencies $\omega_1$, $\omega_2$ is 
\begin{eqnarray} \label{eq:fields}
\vec{\mathcal{B}} & = & \mathcal{B}_z \hat{z} + \vec{\mathcal{B}}_\perp(t) \nonumber \\
\vec{\mathcal{B}}_\perp(t) & = & \mathcal{B}_1 (\hat{x} \ \cos \omega_1 t + \hat{y} \ \sin \omega_1 t) + \mathcal{B}_2 (\hat{x} \ \cos \omega_2 t + \hat{y} \ \sin \omega_2 t) 
\end{eqnarray}

Using the same analysis and notation as the previous section, the energy shift of the ground state is
\begin{equation}
\Delta E_{+1/2} = \frac{1}{4} \Big[ \frac{\gamma^2 s_\perp^2 \mathcal{B}_1 \mathcal{B}_1}{\hbar \omega_1 - \hbar \omega_0} + \frac{\gamma^2 s_\perp^2 \mathcal{B}_2  \mathcal{B}_2}{\hbar \omega_2 - \hbar \omega_0} \Big] + \frac{1}{4} \Big[ \frac{\gamma^2 s_\perp^2 \mathcal{B}_2  \mathcal{B}_1 e^{i (\omega_2 - \omega_1) t}}{\hbar \omega_1 - \hbar \omega_0} + \frac{\gamma^2 s_\perp^2 \mathcal{B}_1  \mathcal{B}_2 e^{-i (\omega_2 - \omega_1) t}}{\hbar \omega_2 - \hbar \omega_0} \Big]
\end{equation}
where we have retained the terms in second order perturbation theory at the beat frequencies between the two Fourier components of the field. The energy shift of the excited state is
\begin{equation}
\Delta E_{-1/2} = \frac{1}{4} \Big[ \frac{\gamma^2 s_\perp^2 \mathcal{B}_1 \mathcal{B}_1 }{-\hbar \omega_1 + \hbar \omega_0} + \frac{\gamma^2 s_\perp^2 \mathcal{B}_2 \mathcal{B}_2 }{-\hbar \omega_2 + \hbar \omega_0} \Big] + \frac{1}{4} \Big[ \frac{\gamma^2 s_\perp^2 \mathcal{B}_1 \mathcal{B}_2  e^{i (\omega_2 - \omega_1) t}}{-\hbar \omega_2 + \hbar \omega_0} + \frac{\gamma^2 s_\perp^2 \mathcal{B}_2 \mathcal{B}_1  e^{-i (\omega_2 - \omega_1) t}}{-\hbar \omega_1 + \hbar \omega_0} \Big].
\end{equation}

Define the difference between the two evolution frequencies to be $\Delta \omega = \omega_2 - \omega_1$. The extra energy difference between the excited and ground states due to the transverse magnetic field is $\Delta E = \Delta E_{+1/2} - \Delta E_{-1/2}$. Up to first order in the evolution frequencies it is given by
\begin{eqnarray}
\Delta E & = & \Big\{- \gamma \frac{ \mathcal{B}_1^2}{2 \mathcal{B}_z} - \gamma \frac{ \mathcal{B}_2^2}{2 \mathcal{B}_z} - \gamma \frac{\mathcal{B}_1 \mathcal{B}_2}{\mathcal{B}_z} \cos \Delta \omega t \Big\} \nonumber \\
& & + \Big[ -\frac{\mathcal{B}_1^2}{2 \mathcal{B}_z^2}\hbar \omega_1 - \frac{\mathcal{B}_2^2}{2 \mathcal{B}_z^2}\hbar \omega_2 - \frac{\mathcal{B}_1 \mathcal{B}_2}{2 \mathcal{B}_z^2} \hbar (\omega_2 + \omega_1) \cos \Delta \omega t \Big] \nonumber \\
& = & \Delta E_{QZ} + \Delta E_g,
\end{eqnarray}
where $\Delta E_{QZ}~(\Delta E_g)$ corresponds to the term in curly (square) brackets. 

All the oscillating terms have been retained above. The phase difference arising from the geometric term is
\begin{equation} \label{eq:gp-2wobble}
\Delta \phi_g = \int_0^T -\Delta E_g \ dt/\hbar = \frac{1}{2} \Big[ \ \frac{\mathcal{B}_1^2}{\mathcal{B}_z^2} \omega_1 T + \frac{\mathcal{B}_2^2}{\mathcal{B}_z^2} \omega_2 T + 
\frac{\mathcal{B}_1 \mathcal{B}_2}{\mathcal{B}_z^2} \frac{(\omega_1 + \omega_2)}{\Delta \omega} \sin \Delta \omega T \ \Big].
\end{equation}
Note that here the geometric phase is not just the sum of phases due to the fields $\mathcal{B}_1$ and $\mathcal{B}_2$, but also contains an oscillating cross term that is proportional to $\sin \Delta \omega T$. This term depends on the phase relationship between the two transverse field components, and cancels (or not) depending on the difference frequency $\Delta \omega$ and the evolution time $T$. This is a physically measurable effect that arises from time-dependent energy shifts that are usually assumed to average away. In the following section, we verify this result by calculating the effect of two evolution frequencies using the usual geometric method. 

\subsection{Geometric approach}

For a spin $S=1/2$ system, the geometric phase difference is $\Delta \phi_g = \Omega_s$. We will evaluate the solid angle $\Omega_s$ enclosed by the tip of the magnetic field vector defined by Equation (\ref{eq:fields}). The solid angle enclosed by a curve on a sphere is given by
\begin{equation}
\Omega_s = \int (1 - \cos \theta) \ d\phi
\end{equation}
where $\theta, \phi$ are the usual spherical polar angles. We use the following definitions
\begin{eqnarray}
\mathcal{B} & = & \sqrt{\mathcal{B}_z^2 + \mathcal{B}_\perp^2} \nonumber \\
\mathcal{B}_z & = & \mathcal{B} \cos \theta 
\end{eqnarray}
and find that the solid angle is 
\begin{eqnarray}
\Omega_s & = & \int \frac{\mathcal{B} - \mathcal{B}_z}{\mathcal{B}} \ d\phi  \nonumber \\
& = & \int \frac{1}{2} \frac{\mathcal{B}_\perp^2}{\mathcal{B}_z^2} \ d\phi \ + \mathcal{O} \Big( \frac{\mathcal{B}_\perp^4}{\mathcal{B}_z^4} \Big) \nonumber \\
& \approx & \frac{\mathcal{A}}{\mathcal{B}_z^2} \label{eq:solidangle}
\end{eqnarray}
where $\mathcal{A} = \int \frac{1}{2} \mathcal{B}_\perp^2 \ d \phi$ is the area enclosed by the transverse field $\vec{\mathcal{B}}_\perp(t)$ in the $xy$-plane. For the purpose of comparison with second order perturbation theory the approximation above, where we retain terms up to second order in $\mathcal{B}_\perp$, is sufficiently accurate.

We can express the area $\mathcal{A}$ in terms of $\vec{\mathcal{B}}_\perp(t), \frac{d \vec{\mathcal{B}}_\perp}{dt}$ and evaluate it for the curve defined in Equation (\ref{eq:fields}):
\begin{eqnarray}
\mathcal{A} & = & \int \frac{1}{2} \mathcal{B}_\perp^2 \ d \phi = \int_0^T \frac{1}{2} \left[ \vec{\mathcal{B}}_\perp(t) \times \frac{d \vec{\mathcal{B}}_\perp}{dt} \right] \cdot \hat{z} \ dt \nonumber \\
& = & \frac{1}{2} \int_0^T dt \ \Big[ (\mathcal{B}_1 \cos \omega_1 t + \mathcal{B}_2 \cos \omega_2 t)(\omega_1 \mathcal{B}_1 \cos \omega_1 t + \omega_2 \mathcal{B}_2 \cos \omega_2 t ) + \nonumber \\
& & (\mathcal{B}_1 \sin \omega_1 t + \mathcal{B}_2 \sin \omega_2 t)(\omega_1 \mathcal{B}_1 \sin \omega_1 t + \omega_2 \mathcal{B}_2 \sin \omega_2 t ) \Big] \nonumber \\
& = & \frac{1}{2} \int_0^T dt \ \Big[ \ \mathcal{B}_1^2 \omega_1 + \mathcal{B}_2^2 \omega_2 + \mathcal{B}_1 \mathcal{B}_2 (\omega_1 + \omega_2) \cos \Delta \omega t \ \Big] \nonumber \\
& = & \frac{1}{2} \Big[ \ \mathcal{B}_1^2 \omega_1 T + \mathcal{B}_2^2 \omega_2 T + \mathcal{B}_1 \mathcal{B}_2 \frac{(\omega_1 + \omega_2)}{\Delta \omega} \sin \Delta \omega T \ \Big].
\end{eqnarray}

Using Equation (\ref{eq:solidangle}), we find that the geometric phase difference is
\begin{eqnarray}
\Delta \phi_g & = & \Omega_s = \frac{\mathcal{A}}{\mathcal{B}_z^2} \nonumber \\
& = & \frac{1}{2} \Big[ \ \frac{\mathcal{B}_1^2}{\mathcal{B}_z^2} \omega_1 T + \frac{\mathcal{B}_2^2}{\mathcal{B}_z^2} \omega_2 T + 
\frac{\mathcal{B}_1 \mathcal{B}_2}{\mathcal{B}_z^2} \frac{(\omega_1 + \omega_2)}{\Delta \omega} \sin \Delta \omega T \ \Big]
\end{eqnarray}
exactly as in the calculation using perturbation theory in Equation (\ref{eq:gp-2wobble}).  We note that Ref. \cite{figure8} discusses a particular example of this situation, where $\omega_1 = 2\omega_2$ and hence the tip of the magnetic field vector executes a figure-8 motion.  

\section{Spin-1 system in an electric field}\label{sec:electric_field}

Using the same formalism as above, the geometric phase can be calculated for a system with more complicated levels such as an atom or a molecule. In this section we will consider the geometric phase for a spin $J=1$ system in an electric field whose direction changes with time. As the electric field vector traces out a loop, the system picks up a geometric phase $\Delta \phi_g = 2 \Omega_s$ between the $|m_J = \pm 1\rangle$ states. We refer the reader to the calculation for this case using the geometric formalism in \cite{bkd}, and calculate the same here using energy shifts. 

Let the evolving electric field be written as 
\begin{eqnarray}
\vec{\mathcal{E}} & = & \mathcal{E}_z \hat{z} + \vec{\mathcal{E}}_{\perp}(t) \\
\vec{\mathcal{E}}_{\perp}(t) & = & \mathcal{E}_{\perp}(\hat{x} \ \cos \omega_\perp t + \hat{y} \ \sin \omega_\perp t) \\
& = & \frac{\mathcal{E}_\perp}{\sqrt{2}} \Big( - \hat{r}_{+1} e^{-i \omega_\perp t} + \hat{r}_{-1} e^{+i \omega_\perp t} \Big)
\end{eqnarray}
where $\hat{r}_{\pm 1} = \mp \frac{(\hat{x} \pm i \hat{y})}{\sqrt{2}}$ are spherical basis vectors.

The Hamiltonian of the system in the electric field is
\begin{eqnarray}
H_{int} & = & -\vec{D} \cdot \vec{\mathcal{E}} \nonumber \\
& = & - D_z \mathcal{E}_z - \frac{D_{-1} \mathcal{E}_\perp e^{-i \omega_{\perp} t} -  D_{+1} \mathcal{E}_\perp e^{i \omega_\perp t}}{\sqrt{2}}.
\end{eqnarray}

\begin{figure}
\includegraphics[width=\columnwidth]{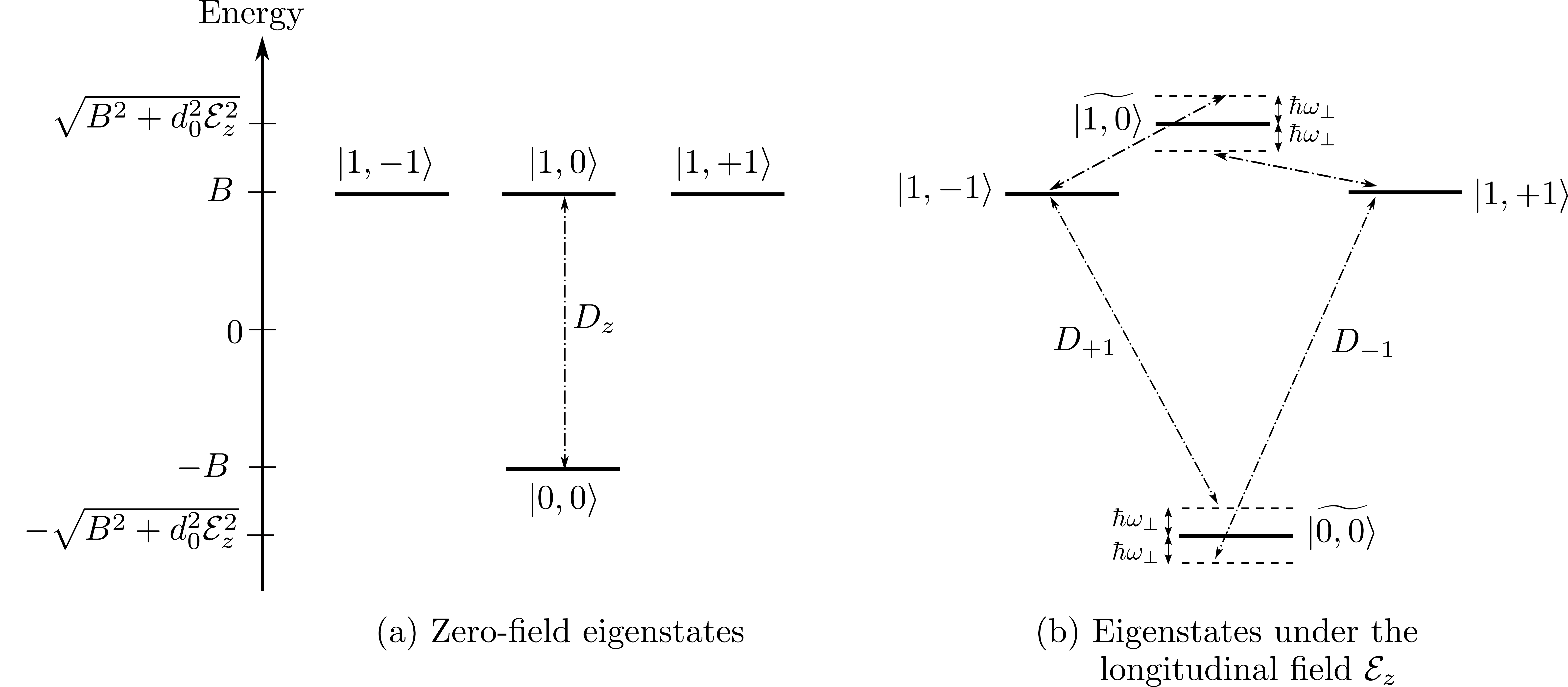}
\caption{Energy levels of a spin $J=1$ system in an electric field. States are labelled by their angular momentum quantum numbers $J, m_J$. Dashed arrows and their labels indicate the states being coupled by the respective components of the dipole moment operator. Under the influence of a longitudinal electric field $\mathcal{E}_z$, the $|0,0\rangle$ and $|1,0\rangle$ states shown in (a) are perturbed into $|\widetilde{0,0}\rangle$ and $|\widetilde{1,0}\rangle$ as shown in (b). The rotating transverse electric field $\vec{\mathcal{E}}_\perp(t)$ induces virtual energy levels (dashed lines) $\hbar \omega_\perp$ above and below the static perturbed states. $\vec{\mathcal{E}}_\perp(t)$ couples these states with the $|1,\pm1\rangle$ states of interest, through the dipole moment operators $D_{\pm1}$.} 
\label{fig:rotor}
\end{figure}

Here we have defined the electric dipole moment operator $\vec{D}$. This operator only couples states of opposite parity, whereas the $|m_J=0, \pm 1\rangle$ sublevels of a $J=1$ level all have the same parity. To calculate the effect of electric fields, it is essential to enlarge the system and include an opposite parity state in addition to the spin-1 sublevels. For simplicity, we consider here the 4-state system consisting of the 3 sublevels of a $J^{\pi} = 1^{-}$ level: $|J=1, m_J = 0\rangle, |J=1, m_J = \pm 1\rangle$, and in addition a $J^{\pi}=0^+$ state: $|J=0,m_J=0\rangle$. Here $J$ denotes the angular momentum and $\pi$ the parity of the state. We refer to states by their $|J,m_J\rangle$ labels from now on. Figure \ref{fig:rotor}(a) shows these states. We are interested in the phase that is picked up between the $|1,\pm 1\rangle$ states due to the evolving electric field. Choose the zero of energy halfway between $|1,0\rangle$ and $|0,0\rangle$, and let the zero-field separation between them be $2B$. The perturbative calculation requires matrix elements of $\vec{D}$, which we write as
\begin{eqnarray}
d_z \hat{z} & = & \langle 1,0|\vec{D}|0,0 \rangle \nonumber \\
- d_{\pm 1} \hat{r}_{\mp 1} & = & \langle 1,\pm 1|\vec{D}|0,0 \rangle.
\end{eqnarray}
Here we have used the Wigner-Eckart theorem to define only the nonzero matrix elements of $\vec{D}$; this can also be used to show that all three of the nonzero matrix elements have a common value: $d_0 \equiv d_z = d_{\pm1}$. 

First, we consider the effect of $\mathcal{E}_z$.  This part of $\vec{\mathcal{E}}$ only mixes the two states $|1,0\rangle$ and $|0,0\rangle$. Under the interaction with this part of the field, the eigenstates are 
\begin{eqnarray}
|\widetilde{1,0}\rangle & = & |1,0\rangle \cos \xi/2 + |0,0\rangle \sin \xi/2 \nonumber \\
|\widetilde{0,0}\rangle & = & |0,0\rangle \cos \xi/2 - |1,0\rangle \sin \xi/2
\end{eqnarray}
where we denote the field-mixed eigenstates with tildes, and the mixing angle is
\begin{eqnarray}
\tan \xi & = & \frac{d_0 \mathcal{E}_z}{B} \nonumber \\
\Rightarrow \sin^2\xi/2 & = & \frac{\sqrt{B^2 + d_0^2 \mathcal{E}_z^2} - B}{2 \sqrt{B^2 + d_0^2 \mathcal{E}_z^2} } \nonumber \\
\cos^2\xi/2 & = & \frac{\sqrt{B^2 + d_0^2 \mathcal{E}_z^2} + B}{2 \sqrt{B^2 + d_0^2 \mathcal{E}_z^2}}.
\end{eqnarray}
The $|\widetilde{1,0}\rangle, |\widetilde{0,0}\rangle$ states have energies $\pm \sqrt{B^2 + d_0^2 \mathcal{E}_z^2}$ respectively. The $|1,\pm 1\rangle$ states remain at their zero-field location $E_{|1,+1\rangle} = E_{|1,-1\rangle} = +B$, as shown in Figure \ref{fig:rotor}(b). The splitting $\Delta_1$ between the $|1,\pm1\rangle$ and $|\widetilde{1,0}\rangle$ levels is usually referred to in the literature as a \emph{tensor Stark} shift \cite{bkd}.

Next we consider the interaction with the rotating transverse field and calculate the effect to second order in perturbation theory. As indicated in Figure \ref{fig:rotor}(b) the transverse field only couples $|1,\pm 1\rangle$ with $|\widetilde{1,0}\rangle$ or $|\widetilde{0,0}\rangle$. Define the following energy denominators for ease of notation
\begin{eqnarray}
\Delta_0 & = & E_{|1,\pm 1\rangle} - E_{|\widetilde{0,0}\rangle} = B + \sqrt{B^2 + d_0^2 \mathcal{E}_z^2} \nonumber \\
\Delta_1 & = & E_{|1,\pm 1\rangle} - E_{|\widetilde{1,0}\rangle} = B - \sqrt{B^2 + d_0^2 \mathcal{E}_z^2}.
\end{eqnarray}
Note that as defined, $\Delta_0 ~ (\Delta_1)$ is positive (negative). The energy shifts $\Delta E_{\pm 1}$ for the $|1,\pm 1\rangle$ states due to the transverse electric field $\mathcal{E}_\perp$ are given by
\begin{eqnarray}
\Delta E_{+1} & = & \frac{1}{2} \Big[ \frac{d_0^2 \mathcal{E}_{\perp}^2 \cos^2\xi/2 }{\Delta_0 + \hbar \omega_\perp} + \frac{d_0^2  \mathcal{E}_{\perp}^2 \sin^2\xi/2}{\Delta_1 + \hbar \omega_\perp} \Big] \nonumber \\
\Delta E_{-1} & = & \frac{1}{2} \Big[ \frac{d_0^2 \mathcal{E}_{\perp}^2 \cos^2\xi/2 }{\Delta_0 - \hbar \omega_\perp} + \frac{d_0^2 \mathcal{E}_{\perp}^2 \sin^2\xi/2 }{\Delta_1 - \hbar \omega_\perp} \Big].
\end{eqnarray}

Where the $|1,\pm 1\rangle$ states were degenerate before, they are now split by the energy $\Delta E$ given by
\begin{eqnarray}
\Delta E & = & \Delta E_{+1} - \Delta E_{-1} \nonumber \\
& = & - \frac{d_0^2 \mathcal{E}_{\perp}^2 \cos^2\xi/2}{\Delta_0^2} \hbar \omega_\perp -  \frac{d_0^2 \mathcal{E}_{\perp}^2 \sin^2\xi/2}{\Delta_1^2} \hbar \omega_\perp + \mathcal{O}(\omega_\perp^2).
\end{eqnarray}
We have again retained only the terms to least order in $\omega_\perp$ to illustrate the adiabatic part of the phase. Substituting the values of $\xi, \Delta_0, \Delta_1$ we get
\begin{eqnarray}
\Delta E & = & - d_0^2 \mathcal{E}_\perp^2 \Bigg[\frac{\sqrt{B^2 + d_0^2 \mathcal{E}_z^2} + B}{2 \sqrt{B^2 + d_0^2 \mathcal{E}_z^2} } \ \frac{1}{\Big(\sqrt{B^2 + d_0^2 \mathcal{E}_z^2} + B\Big)^2}  + \ \nonumber \\
& & \frac{\sqrt{B^2 + d_0^2 \mathcal{E}_z^2} - B}{2 \sqrt{B^2 + d_0^2 \mathcal{E}_z^2} } \ \frac{1}{\Big(\sqrt{B^2 + d_0^2 \mathcal{E}_z^2} - B\Big)^2} \Bigg] \hbar \omega_\perp \\
& = & - \frac{\mathcal{E}_\perp^2}{\mathcal{E}_z^2} \hbar \omega_\perp  \ = \Delta E_g.
\end{eqnarray}

Compared to the case with a magnetic field, in a pure electric field the only energy difference between $|1,\pm 1\rangle$ is the geometric contribution $\Delta E_g$. The relative phase between $|1,\pm 1\rangle$ after a time duration corresponding to a single complete cycle of evolution, $T = 2 \pi/\omega_\perp$,  is
\begin{eqnarray}
\Delta \phi_g & = & -\Delta E_g T/\hbar = \frac{\mathcal{E}_\perp^2}{\mathcal{E}_z^2} \omega_\perp T \nonumber \\
& = & 2 \pi \frac{\mathcal{E}_\perp^2}{\mathcal{E}_z^2} = 2 \Omega_s
\end{eqnarray}
and the phase shift again has a simple geometric interpretation in terms of the solid angle enclosed by the evolving electric field.

\section{Combined effect of electric and magnetic fields}\label{sec:EplusB}

The formalism developed in the previous section can be easily extended to a more complicated case, where the system interacts with both electric and magnetic fields. The standard geometric approach in this case requires us to calculate all instantaneous eigenvectors of the system, and their gradients with respect to the 6-dimensional parameter space that describes the Hamiltonian and its evolution (3 degrees of freedom each for the electric and magnetic fields). Energy shifts on the other hand can be calculated in a straightforward manner, as we shall see in the following. 

\subsection{Rotating electric field, static magnetic field}
We will first examine the same spin-1 system as in the previous section and understand what happens to it when a static magnetic field $\vec{\mathcal{B}} = \mathcal{B}_z \hat{z}$ is imposed along with the revolving electric field. The interaction Hamiltonian of the system is now
\begin{eqnarray}
H_{int} & = & -\vec{D} \cdot \vec{\mathcal{E}}  - \vec{\mu} \cdot \vec{\mathcal{B}} \nonumber \\
& = & - D_z \mathcal{E}_z - \mu_z \mathcal{B}_z - \frac{D_{-1} \mathcal{E}_\perp e^{-i \omega_{\perp} t} -  D_{+1} \mathcal{E}_\perp e^{i \omega_\perp t}}{\sqrt{2}}.
\end{eqnarray}
with $\vec{\mu}$ the magnetic moment of the $J=1$ state. 

\begin{figure}
\includegraphics[width=\columnwidth]{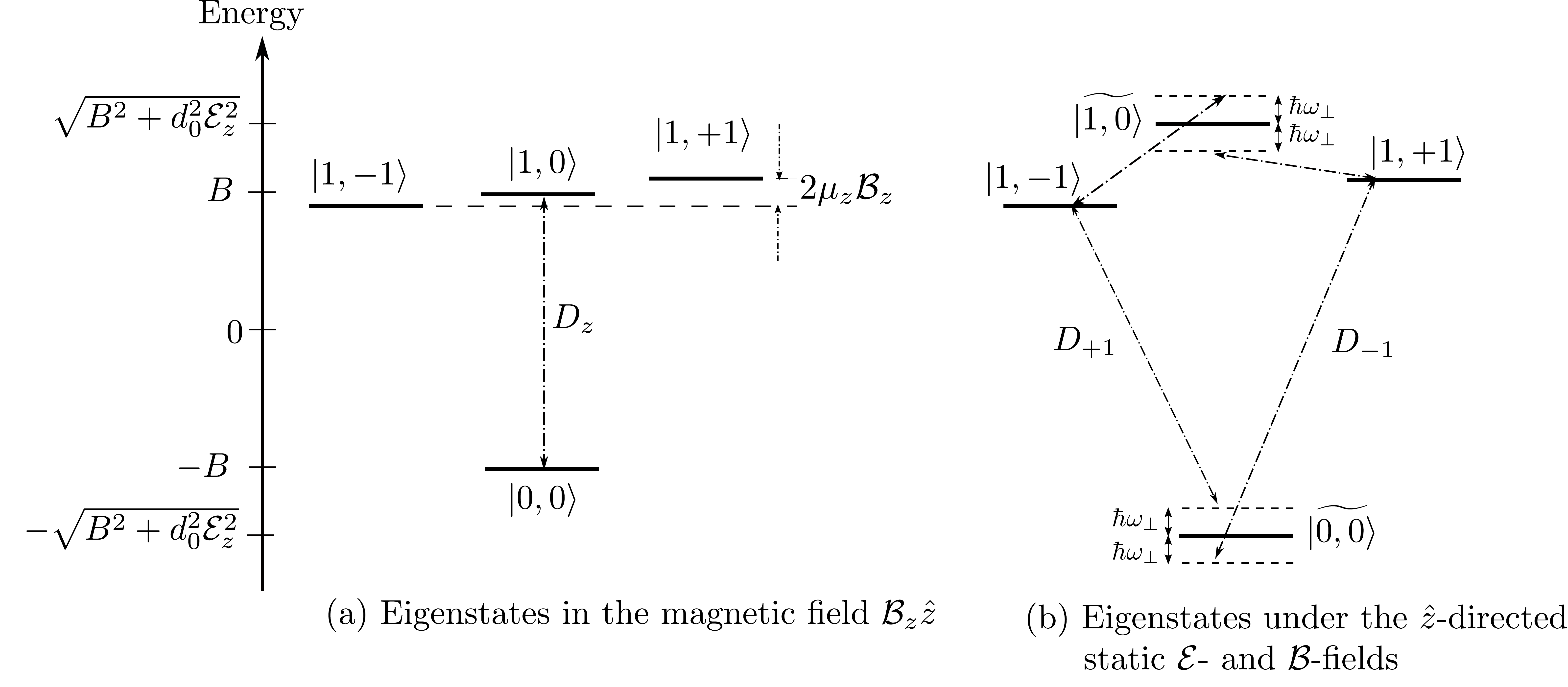}
\caption{Energy levels of a spin $J=1$ system in a combination of electric and magnetic fields.  The state labels and symbols are the same as in Figure \ref{fig:rotor}. In (b), the figure as shown corresponds to a case where the Zeeman shift is somewhat smaller than the tensor Stark shift: $\mu_z \mathcal{B}_z \lesssim \Delta_1$.}
\label{fig:bfield-rotor}
\end{figure}

To illustrate the basic ideas we again use a perturbative approach, to study the influence of the transverse time-dependent components of $\mathcal{E}$ on the eigenstates in the presence of only $\mathcal{E}_z$ and $\mathcal{B}_z$. The eigenstates of the system are shown in Figure \ref{fig:bfield-rotor}. We will carry over the notation and definitions from the previous section. 

The essential effect of the $\mathcal{B}_z$ field is in the energy denominators that appear in the  AC Stark shift. The energy shifts of the $|1,\pm 1\rangle$ states (in addition to their Zeeman shifts in the magnetic field) are given to lowest order in $\mathcal{E}_\perp$ by
\begin{eqnarray} \label{eq:BE}
\Delta E_{+1} & = & \frac{1}{2} \Big[ \frac{d_0^2 \mathcal{E}_{\perp}^2 \cos^2\xi/2 }{\Delta_0 + \mu_z \mathcal{B}_z + \hbar \omega_\perp} + \frac{d_0^2  \mathcal{E}_{\perp}^2 \sin^2\xi/2}{\Delta_1 + \mu_z \mathcal{B}_z + \hbar \omega_\perp} \Big] \nonumber \\
\Delta E_{-1} & = & \frac{1}{2} \Big[ \frac{d_0^2 \mathcal{E}_{\perp}^2 \cos^2\xi/2 }{\Delta_0 - \mu_z \mathcal{B}_z - \hbar \omega_\perp} + \frac{d_0^2 \mathcal{E}_{\perp}^2 \sin^2\xi/2 }{\Delta_1 - \mu_z \mathcal{B}_z - \hbar \omega_\perp} \Big].
\end{eqnarray}

To make a connection with the adiabatic limit, we expand the energy shift $\Delta E$ between $|1,\pm1\rangle$ up to first order in $\hbar \omega_\perp$, and write 
\begin{eqnarray}
\Delta E = \Delta E_{+1} - \Delta E_{-1} \approx \mathcal{C} + \mathcal{D} \ \hbar \omega_\perp.
\end{eqnarray}

We interpret $\mathcal{C}$ as the analog of a (static) quadratic Zeeman/Stark shift due to the inclusion of $\mathcal{E}_\perp$, and $\Delta E_g = \mathcal{D} \ \hbar \omega_\perp$ as the energy shift that leads to a geometric phase difference $\Delta\phi_g = -\mathcal{D} \omega_\perp T$. Table \ref{tab:1} shows the values of the geometric phase $\Delta \phi_g$ in some limiting cases, corresponding to regimes where one of the various energy scales in the problem is dominant. The relevant energy scales are the Zeeman shift $\mu_z \mathcal{B}_z$, the Stark matrix element $d_0 \mathcal{E}_z$, the Stark shift of the $|1,0\rangle$ state $|\Delta_{1}|$, and the zero-field energy splitting $2B$ between $J^\pi=0^+$ and $J^\pi=1^-$ levels.

Note that in general the geometric phase cannot be related to an obvious solid angle, due to the complicated parameter space of the Hamiltonian. While the evaluation of the actual solid angle by geometric means can be rather arduous, the geometric phase in every case can be simply calculated from the energy shift between $|1,\pm 1\rangle$. This energy shift can be computed perturbatively, or by numerically diagonalizing the Hamiltonian if necessary. An important feature to note in Table \ref{tab:1} is that the zero-field energy splitting between $J^\pi = 0^+$ and $J^\pi=1^-$ levels appears irreducibly in some of the limiting expressions. This shows that the inclusion of the opposite parity (\emph{e.g.} $J^\pi=0^+$) state is necessary to properly describe geometric phases in a combination of electric and magnetic fields.

\begin{table}[htb!]
  \begin{center}
    \begin{tabular}{lll}
		\hline
		Case & ~~Limiting condition &  ~~~ Geometric phase difference, $\Delta\phi_g$	\\
		\hline
		~ \\
		~I & $\mu_z \mathcal{B}_z \ll |\Delta_1|, d_0 \mathcal{E}_z$  & ~ ~ ~ $\displaystyle \Big(\frac{\mathcal{E}_\perp}{\mathcal{E}_z}\Big)^2 \omega_\perp T$ \\
		~ \\
		~II & $|\Delta_1| \ll \mu_z \mathcal{B}_z \ll d_0 \mathcal{E}_z$ & ~ ~ ~ $\displaystyle \Big(\frac{\mathcal{E}_\perp}{\mathcal{E}_z} \Big)^2 \Big(\frac{\Delta_1}{\mu_z \mathcal{B}_z}\Big)^2 \omega_\perp T$ \\
		~ \\
		~III & $d_0 \mathcal{E}_z \ll \mu_z \mathcal{B}_z \ll 2B$ & ~ ~ ~ $\displaystyle \Big(\frac{ \mathcal{E}_\perp}{\mathcal{E}_z}\Big)^2 \Big(\frac{d_0\mathcal{E}_z}{2B}\Big)^2 \omega_\perp T$ \\
		~ \\
		~IV & $d_0 \mathcal{E}_z, 2B \ll \mu_z \mathcal{B}_z$ & ~ ~ ~ $\displaystyle \Big(\frac{ \mathcal{E}_\perp}{\mathcal{E}_z}\Big)^2 \Big(\frac{d_0 \mathcal{E}_z}{\mu_z \mathcal{B}_z}\Big)^2 \omega_\perp T$ \\
		~\\
		\hline
		\end{tabular}
\caption{Analytical expressions in some limiting conditions, for the geometric phase in a static magnetic field $\mathcal{B}_z$ and electric field $\mathcal{E}_z$, along with a rotating transverse electric field component $\mathcal{E}_\perp$. In each case, the geometric phase between $|1,\pm 1\rangle$ up to first order in $\omega_\perp$ is shown.} \label{tab:1}  
  \end{center}
\end{table}

\subsection{Rotating magnetic field, static electric field}

As another illustration of the energy shift formalism in a case with combined electric and magnetic fields, consider the case where the system experiences a static $\hat{z}$-directed electric field, and a revolving magnetic field $\vec{\mathcal{B}} = \mathcal{B}_z \hat{z} + \mathcal{B}_\perp (\hat{x} \ \cos \omega_\perp t + \hat{y} \ \sin \omega_\perp t)$. The interaction Hamiltonian is
\begin{eqnarray}
H_{int} & = & -\vec{D} \cdot \vec{\mathcal{E}}  - \vec{\mu} \cdot \vec{\mathcal{B}} \nonumber \\
& = & - D_z \mathcal{E}_z - \mu_z \mathcal{B}_z - \frac{\mu_{-1} \mathcal{B}_\perp e^{-i \omega_{\perp} t} -  \mu_{+1} \mathcal{B}_\perp e^{i \omega_\perp t}}{\sqrt{2}}.
\end{eqnarray}
where the spherical components of the magnetic moment $\vec{\mu}$ are defined as follows.
\begin{eqnarray}
\mu_z \hat{z} & = & \langle 1,1|\vec{\mu}|1,1 \rangle \nonumber \\
- \mu_{\pm 1} \hat{r}_{\mp 1} & = & \langle 1,\pm 1|\vec{\mu}|1,0 \rangle.
\end{eqnarray}
Once again, the Wigner-Eckart theorem has been used to define the non-zero matrix elements (which have a common value): $\mu_0 \equiv \mu_z = \mu_{\pm 1}$. 
 
The energy shifts are
\begin{eqnarray}\label{eq:EplusB}
\Delta E_{+1} & = & \frac{1}{2} \Big[ \frac{\mu_0^2 \mathcal{B}_{\perp}^2 \sin^2\xi/2 }{\Delta_0 + \mu_z \mathcal{B}_z + \hbar \omega_\perp} + \frac{\mu_0^2  \mathcal{B}_{\perp}^2 \cos^2\xi/2}{\Delta_1 + \mu_z \mathcal{B}_z + \hbar \omega_\perp} \Big] \nonumber \\
\Delta E_{-1} & = & \frac{1}{2} \Big[ \frac{\mu_0^2 \mathcal{B}_{\perp}^2 \sin^2\xi/2 }{\Delta_0 - \mu_z \mathcal{B}_z - \hbar \omega_\perp} + \frac{\mu_0^2 \mathcal{B}_{\perp}^2 \cos^2\xi/2 }{\Delta_1 - \mu_z \mathcal{B}_z - \hbar \omega_\perp} \Big].
\end{eqnarray}

As in the previous section, we expand the energy shift $\Delta E$ between $|1,\pm1\rangle$ up to first order in $\hbar \omega_\perp$ 
\begin{eqnarray}
\Delta E = \Delta E_{+1} - \Delta E_{-1} \approx \mathcal{C} + \mathcal{D} \ \hbar \omega_\perp 
\end{eqnarray}
and focus our attention on the relative energy shift $\Delta E_g = \mathcal{D} \ \hbar \omega_\perp$ that leads to a geometric phase $\Delta \phi_g$. The relevant energy scales in the problem are the same as in the previous section. We consider cases where $\Delta_1 \ll 2B$ \emph{i.e.} the system is weakly electrically polarized by the $\mathcal{E}_z$-field. Table \ref{tab:2} lists the geometric phase in some limiting cases where the analytic expression becomes simple. 

As expected, when the Zeeman splitting is sufficiently large (Cases I and II) the geometric phase reduces to the result expected in the pure magnetic field scenario. Case III is interesting and quite relevant to real world experiments. Compared to the situation in a pure magnetic field, the geometric phase in the presence of a strong tensor Stark splitting is significantly suppressed, by a factor $\displaystyle \Big(\frac{\mu_z\mathcal{B}_z}{\Delta_1}\Big)^2$ which could be $\sim 10^{-6}$ in experiments using polar molecules to search for the electron electric dipole moment \cite{hinds, cornellprivate,gfactor_difference}.

The expressions listed in Table \ref{tab:2} are special cases derived from the more general Equation (\ref{eq:EplusB}) for the energy shift between $|1,\pm1\rangle$. \emph{A priori} it is not intuitively evident how Equation (\ref{eq:EplusB}) or the limiting cases in Table \ref{tab:2} can be related to a solid angle in the parameter space of the Hamiltonian. But reformulating the problem in terms of energy level shifts allows us to exchange a convoluted geometric construction for a simple algebraic calculation.

\begin{table}[htb!]
  \begin{center}
    \begin{tabular}{lll}
		\hline
		Case & ~~Limiting condition &  ~~~ Geometric phase difference, $\Delta \phi_g$	\\
		\hline
		~ \\
		~I & $|\Delta_1| \ll 2 B \ll \mu_z \mathcal{B}_z$  & ~ ~ ~ $\displaystyle \Big(\frac{\mathcal{B}_\perp}{\mathcal{B}_z}\Big)^2 \omega_\perp T$ \\
		~ \\
		~II & $|\Delta_1| \ll \mu_z \mathcal{B}_z \ll 2 B$  & ~ ~ ~ $\displaystyle \Big(\frac{\mathcal{B}_\perp}{\mathcal{B}_z}\Big)^2 \omega_\perp T$ \\
		~ \\
		
		~II & $\mu_z \mathcal{B}_z \ll |\Delta_1| \ll 2B$ & ~ ~ ~ $\displaystyle \Big(\frac{\mu_0 \mathcal{B}_\perp}{\Delta_1}\Big)^2 \omega_\perp T$ \\
		~\\
		\hline
		\end{tabular}
\caption{Analytical expressions in some limiting conditions, for the geometric phase in a static electric field $\mathcal{E}_z$ with a rotating transverse magnetic field component $\mathcal{B}_\perp$. In each case, the geometric phase between $|1,\pm 1\rangle$ up to first order in $\omega_\perp$ is shown.} \label{tab:2}  
  \end{center}
\end{table}

\section{Summary}
We have worked out some simple examples where the geometric phase arises due to magnetic or electric fields whose direction changes over time. Identifying the geometric phase as nothing other than the phase due to off-resonant energy shifts enables the use of textbook quantum mechanics to calculate it in every case. In contrast to the usual geometric approach, geometric phases due to non-adiabatic evolution and/or non-cyclic paths in parameter space can be calculated in a straightforward way using energy shifts. Furthermore, the energy shift approach is convenient for understanding geometric phases when complex level structures or multi-dimensional parameter spaces are involved. Instead of tracking all the eigenstates as the Hamiltonian evolves along a path in its parameter space, the problem is reduced to one of simple and intuitive algebra. This simplification enables the precise calculation of geometric phase effects in experimentally relevant situations, where the evolution of the fields experienced by atoms or molecules can be quite complicated. Under conditions where the transverse fields are sufficiently small, using perturbation theory to calculate energy shifts (as described here) gives particularly simple expressions even when the relation to the actual geometry of the situation is unclear.  

The analysis in terms of energy shifts outlined here is not meant to replace the geometric approach. The beautiful and general results known from geometric arguments emphasize some of the topological aspects of geometric phases, which cannot be easily obtained in the energy shift formulation. However, it is useful to realize that the full machinery of the geometric formalism is not in general necessary to calculate the relevant phases.  Moreover, from the pedagogical point of view, we hope that recasting the geometric phase in terms of familiar concepts (phase differences due to energy shifts) may make it seem a little less mysterious.

\section*{Acknowledgments}
We thank Steve Lamoreaux and Steve Girvin for interesting discussions. This work was supported by the National Science Foundation grant	PHY0555462.

\bibliography{geomphase}

\end{document}